# EXPERIENCE ON RE-ENGINEERING APPLYING WITH SOFTWARE PRODUCT LINE

Waraporn Jirapanthong

Faculty of Information Technology, Dhurakij Pundit University, Bangkok, Thailand
waraporn.jir@dpu.ac.th

**Abstract.** In this paper, we present our experience based on a reengineering project. The software project is to re-engineer the original system of a company to answer the new requirements and changed business functions. Reengineering is a process that involves not only the software system, but also underlying business model. Particularly, the new business model is designed along with new technologies to support the new system. This paper presents our experience that applies with software product line approach to develop the new system supporting original business functions and new ones.

## 1  Introduction

It becomes so common that businesses are driven by software systems. In particular, the companies have been relied on performing of software systems. The software systems have been developed and evolved over years. Some of those systems have been gradually evolved and become very complicated systems. This is because of the change of original business functions and adding of new ones. Some software systems have been developed over the last 20 years with a very high level of success. However, the software systems now have two significant issues. Firstly, the original business model is now changed. Secondly, the factor of new technologies leads a change. In particular, business functions are needed to response the new architectural techniques and platforms which appear more attractive and some less economical. An idea of re-engineering is definitely very important and has been ongoing to draw the attention of the research community.

Additionally, software product line is originally introduced to serve the reuse practice in an organization having a large number of products, which drives issues such as highly expensive, complex, and tedious tasks. The idea of product line was motivated by the need to systematize a number of products more effectively and the fact that these products have a certain set of common and special functionalities. For example, a mobile-phone company has created a mobile-phone family that contains a set of mobile-phones. Some lower end mobile-phones have similar basic functionalities but different hardware capacities to offer competitive price. Region-based mobile-phones are designed for different transmission and signaling standards, depending on regional diversity; thereby, the company provides different functionalities of mobile phones to support different regions.

In this paper we present our experience based on a re-engineering project. The project is to reengineer the original system to response changes. The new system is designed and implemented by applying with software product line approach. Moreover, the system prototyping technique is used to present the new system. The first section has been an introduction of this work. The second section describe the background of this work including software product line, and the original and new systems. In the next section we present our methodology. The following section, we describe the evaluation of work. The final section, we present the conclusions and discussion.

## 2  Background

In this section, the framework of software product line artefacts as well as the original and new systems are presented.

### 2.1  Software Product line

Software product line systems share a common set of features and are developed based on the reuse of core assets. Many approaches [1],[2],[3],[4],[5],[6],[7],[8], [9],[10],[11],[12],[13],[14],[15],[16],[17] have been proposed to support software product line development. Those approaches mainly focus on domain engineering which is concerned the identification and analysis of commonality and variability principles among applications in a domain in order to engineer reusable and adaptable components and, therefore, support product line development. There are three steps for domain engineering: (a) domain analysis is the process of identifying, collecting, organizing and representing the relevant information in a domain, based upon the study of existing systems and their developing histories, knowledge captured from domain experts, underlying theory, and emerging technology within a domain [14]. Software artefacts that are produced during the activity of domain analysis are called reference requirements, which define the products and their requirements in a family. The reference requirements contain commonality and variability of the product family. The activities occur during the domain analysis are scoping, defining of commonality and variability, and planning for product members and features.

(b) domain design is the process of developing a design model from the products of domain analysis and the knowledge gained from the study of software requirements or design reuse and generic architectures [8]. Software artefacts that are produced during the activity of domain design are called software product line architecture, which forms the backbone of integrating software systems and consists of a set of decisions and interfaces which connect software components together.

Software product line architecture differs from an architecture of individual systems that it must represent the common design for all product members and variable design for specific product members. The activities occur during the domain design are defining and evaluation of software product line architecture.

(c) domain implementation is the process of identifying reusable components based on the domain model and generic architecture [5]. Software artefacts that are produced during the activity of domain implementation are called reusable software components. The activity is focused on the creation of reusable software components e.g. source codes and linking libraries that are later assembled for product members At the end of the domain engineering process, an organization is ready for developing product members.

Additionally, application engineering is a systematic process for the creation of a product member from the core assets created during the domain engineering. Domain engineering assures that the activities of analysis, design and implementation of a product family are thoroughly performed for all product members, while application engineering assures the reuse of the core assets of the product family for the creation of product members. There are activities such as: (i) requirements engineering, which is a process that consists of requirements elicitation, analysis, specification, verification, and management; (ii) design analysis, which is a process that is concerned with how the system functionality is to be provided by the different components of the system; and (iii) integration and testing, which is a process of taking reusable components then putting them together to build a complete system, and of testing if the system is working appropriately.

However, although the support for identifying and analysing common and variable aspects among applications and the engineering of reusable and adaptable components are important for software product line development, they are not easy tasks. This is mainly due to the large number and heterogeneity of documents generated during the development of product line systems.

### 2.2 The Original and New Systems

The original system is a support system for inventory management of all the regional companies. Some business functions of the original system are performed manually and some are supported automatically. For the business functions which are manually performed, they are driven by clerks who are in charge those business functions for a long time. Basically, the head company spends a large portion of capital investment (over $0.1 Million per month) to perform regular business functions.

All of the business functions have now designed and implemented as a new business model. The expected benefits in terms of economic reasons are that (i) the company wanted to decrease the capital investment of performing regular business functions, (ii) the obsolete machines and equipment are needed to be eliminated and replaced with high-end machines and equipment e.g. database management systems, personal computers, network equipment. The new system is expected to improve performance and reduce human errors. To do so, the new system is driven based on an automatic environment. For example, originally, users are human operators (clerks).

Presently, a clerk works with the new system which is a web-based application as a part of mechanism. Originally, forms are simple interfaces allowing users to enter parameters such as text or numerical fields and then submit them to the system. The new forms interface running on web-based applications is optimized for parameter entry rather than for performing semi-automatic complicated tasks requiring more complex user interaction. The architecture of the old system has problems due to its size, platform and complexity. It was developed and has evolved in the last twenty years. This leads the system grew very large. In addition,, it is a closed system. It is not an easy way to modularize it and replace any of its parts. This is because there are no clearly defined interfaces or logical modules. Moreover, the original system is based on a hierarchical database running on a mainframe which now appears very unattractive. Its difficulties i.e. usability, size, platform, and performance become problems to business functions nowadays.

## 3  Our Methodology

### 3.1 Re-Engineering Activities

We considered the technique of software prototyping due to its several benefits i.e. getting direct-feedback from the users in early stage of the software project development, a team developer can insight into the precision of software specification comparing with the software program. We then believe that building a prototype of reengineering system would lead to a complete reengineering system less effortlessly. The prototype is designed and implemented to represent a new system which preserves significant set of old functionalities. The original system was based on performing within the constraints of transaction load whereas the prototype of the new system presents carrying out in different approach. However, the approach serves those purposes of the original system.

   Firstly, the prototype shows the system migration by driving a mainframe-based system to more open systems due to its motivation i.e. cost savings. Many tools are applied to support automatically converting the data from one form to another and the code from one platform to another. In particular, the prototype show the new system architecture based on client-server computing and relational database to the application domain.

   Secondly, the prototype shows that the new system allows us to derive non-functional requirements (performance-based) and to scale up various relational database systems. Finally, the prototype provides various user interfaces that enable stakeholders to experiment and meet the complete requirements for the final software product.

   However, there are significant differences between the old and new business model. Therefore, the new systems need to accommodate the old preserved requirements and the new one. For example, many tasks which are used to perform by multiple clerks of the old system would be replaced by the new system. There are only empowered clerks who can handle complicated tasks are still needed to remain in the system. Most of business functions are expected to perform in an automatic way; however, not for exception handling.

As mentioned earlier, the new system relies on a three level client-server architecture. A user directly interacts with the new system via a web-based application as a front-end, and the front-end talks to the business function through RPC protocol running over TCP/IP. In this point, the business function may be serviced by another server under the web-service technology. The back-end is also responsible to inquiries of RDBMS and to handle multiple instances of user accesses. In our approach, we also apply XML technology to support multiple accesses to RDBMS and multi-users.

In addition to the development of the new system, we decided to apply a software product line approach as a software development methodology. The whole system is analysed and designed with the concern of software product line. This is believed that it would take more benefits when the system needs to evolve or response the change in the future. It is also believed that the software product would be produced within the time constraint.

### 3.2 Software Development Team and Projects

This activity was conducted to identify the practices that clearly contribute to software project success. It investigated team knowledge, allocated resources, and deployed software processes. We established a software development team with the equivalent skills to be responsible the projects. The software development team that has experience in software development was participated in the study. The software development team was established and included a system analyst, a project manager, and four software developers. This research conducted an experiment involving many small software development projects that are divided from the new system. The projects have similar requirements and some different requirements. Three software projects were designed in order to narrow set of requirements and those are based on business functions.

Before starting the projects, all developers were trained regarding software product line processes and techniques. These developers were then tested for their understanding of software product line practices by using questionnaires. Those who passed the test were assumed to be ready to implement projects using software product line. At the beginning of a project the developers need to take several days to envision the high-level requirements and to understand the scope of the release. The goal of this activity is to find what the project is all about, not to document in detail. The developers then started developing a set of software projects by following the software product line practices. They studied and analyzed all projects together and produced the software artefacts: (i) reference requirements; (ii) software product line architecture; and (iii) software components.

The artefacts were checked before submitting to the domain repository to be ready for application engineering process. Next, three software products were created based on the domain artefacts (i.e. reference requirements, software product line architecture, and software components). Before the software was accepted by stakeholders, we ran test cases on the software. When the software passed all test cases, the projects are completed. The whole software product line process is shown in Figure 1. We then calculated and analyzed the qualitative and quantitative aspects of domain engineering process and application engineering process for each project. Then we checked the developers conform to software product line practices

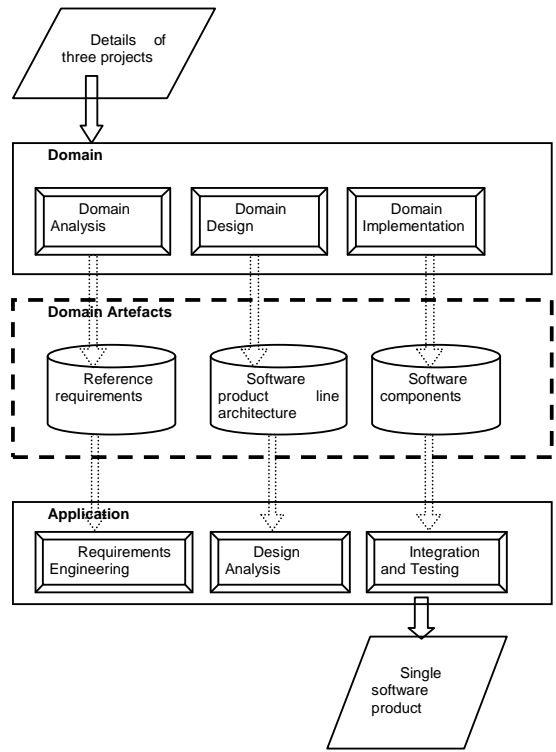

**Fig. 1.** Software Product Line Process.

## 4 Evaluations

In this section, we analyse and evaluate the experiments by focusing on two aspects of measurement: (a) Qualitative and  (b) Quantitative measurement.

**4.1 Qualitative Measurement**

We evaluate our work of re-engineering which applies with software product line methodology. We focus on the performance analysis of software development team. We measured the time spending during the development. First of all, we measured the time spending on the development of core assets i.e. reference requirements, software product line architecture, and reusable software components. We also measured the business functions created in the new system. We then outlined the matrix of business functions created and developing time spent in order to measure the average time on development of each business function including the average time of the core asset development.

Moreover, we wanted to measure the performance of service may be corrupted due to more user accesses. At first, we expected the new system could perform based on the assumption of small queue length. Otherwise, with the response time to a user that is acceptable is less than 1 second; the system can only support 8-15 users.

In conclusion, the experiments showed the performance of development activities very well in terms of response time and capacities. The software product line methodology and our particular hardware platform were good enough to serve a number of developers and software artefacts.

In addition, in general, qualitative methods and tools for system analysis can address the problem of how to empirically determine the context of software process. In this research, we focused on comparison between two software process methodologies how they are practiced. As mentioned, we have conducted the survey and interview. It has been observed that the users are satisfied with the software product line resulting projects and teamwork. Moreover, the software product line developers satisfied the process that emphasis the software more than the documentation. However, it has been also noticed that it is easier to train waterfall-based practices to inexperience developers but some experience developers tend to resist some software product line practices because (a) they have to change their style in working, and (b) it costs them for establishing the software product line artefacts.

According to the survey, it is found that 33% of developers tend to resist software product line practices with the above reasons, whereas 70% of developers are positive to using software product line practices. Particularly, 82% of developers are satisfied when performed the maintenance phase with software product line. Some of software product line artefacts are used during the maintenance phase. And it is satisfied by the developers. However, application engineering process depends on developer' skill. Moreover, the waterfall-based developers are unsatisfied to frequently update the documentation.

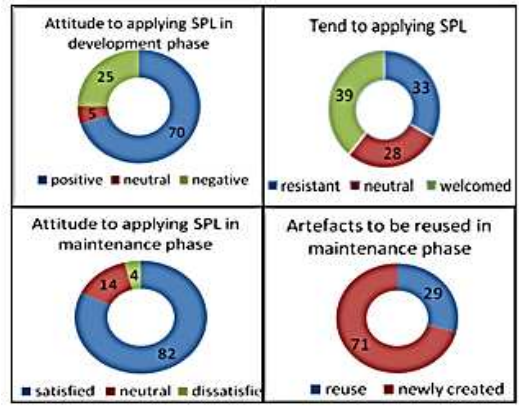

**Fig. 2.** Qualitative Measurement

### 4.2 Quantitative Measurement

Basically quantitative metrics are fundamentally limited to the measurement of the size of system, time and effort spent during software development process. In this research, we measured the total of work hour spent during development and maintenance phases as well as the errors during the phases by following the software processes. In particular, we take account into the number of items causing a software system false. As mentioned earlier, the developer team was required to develop a set of software products by applying with software product line process.

As shown in Table 1, the result shows the effort of software product line-based projects. Comparing the old projects, software product line-based projects enhance the productivity by using existing software artefacts. The methodology supports software reuse at the largest level of granularity. The more software artefacts are reused, the less time is spent. Although, developers spent extra time and effort to establish domain artefacts, it seems the trend of effort for new products in the same product line would decrease. Moreover, we learnt that user are involved at the inception of project determined requirements and contractual agreement. Developers wrote all documents (e.g. design documents) before coding. Then stakeholders changed some requirements, maybe after they acquired finally product, developers needed to significantly redesign and edit their documents. This took a lot of effort to achieve the task.

**Table 1** The Effort and Errors During Development Phase

|  | Product Name | Work-hour | Error |
|---|---|---|---|
| Domain Engineering | - | 620 | 22 |
| SPL-based project1 | PL_1 | 315 | 17 |
| SPL-based project2 | PL_2 | 240 | 15 |
| SPL-based project3 | PL_3 | 215 | 15 |

However, the number of errors which occur during the development phase of software product line is high. Also, some defects are discovered during the integration process for a product member. It took some effort to fix them.

All projects were completed by a software development team in 6 months with an effort of 6 person-months. The development time included understanding of requirements, design, implementation, and testing. The final system includes 17 user interfaces for all different types of users. There is a limitation of 4 simultaneous connections from one computer to one server; otherwise, ther is no limitation of the quantity of connections from one computer to many servers. Specifically, if the server is upgraded to a faster system, it would support more requests with shorter time.

## 5 Conclusion and Discussion

The prototype was designed and created in order to be applied as a tool to drive the reengineering activities. We learnt a lot of issues during the reengineering activities i.e. the development of the new system and the migration of one. Firstly, performing business functions in an automatic way is a highly desirable feature. Secondly, technically, user interfaces representing business functions should be based on real world in order to support end-users. In fact, the design and implementation of user interfaces is not an easy task. Next, the new system provides better environments to facilitate their process. According to the performance requirement which includes the expectation on speed of data access should be reflected naturally. However, the combination of different techniques for building re-engineered systems is a very difficult task. Next, although the automation of business functions is highly desirable, many clerks of the old system that are replaced are not satisfied with the new solution.

Moreover, the results show that the effort metric of software product line-based projects is less than single software projects. This is compared with the experiences from the development of old system. Software product line-based projects enhance the productivity by using existing software artefacts. The methodology supports software reuse at the largest level of granularity. However, developers spent time and effort to establish domain artefacts. Also, some defects are discovered during the integration process for a product member. It took some effort to fix them. Developers wrote all documents before coding. Then customers changed some requirements, maybe after they acquired finally product, developers needed to significantly redesign and edit their documents. This took a lot of effort to achieve the task.

However, software product line is unsuitable for all projects. It serves the reuse practice in an organization having a large number of products, which have similar requirements and some differences. Developers must consider the characteristics of the project to ensure software product line is appropriate. In the other hand, conventional software development approaches i.e. waterfall approach may be suitable to serve a software project which is small and has solid requirements. Also, the developers are responsible for estimating the effort required to implement the requirements which they will work on. Although the developers may not have the requisite estimating skills, it does not take long for them to get better at estimating when they know and get familiar with the software process methods.